\documentclass[conference]{IEEEtran}
\ifCLASSINFOpdf
\else
  \usepackage[dvips]{graphicx}
\fi
\usepackage{color}
\usepackage{dblfloatfix}

\begin{document}
%
\title{Multiplexed readout demonstration of a TES-based detector array in a resistance locked loop}

\author{\IEEEauthorblockN{Jan van der Kuur\IEEEauthorrefmark{1},
Luciano Gottardi\IEEEauthorrefmark{1},
Mikko Kiviranta\IEEEauthorrefmark{2}, 
Hiroki Akamatsu\IEEEauthorrefmark{1},
Pourya Khosropanah\IEEEauthorrefmark{1},\\
Roland den Hartog\IEEEauthorrefmark{1},
Toyoaki Suzuki\IEEEauthorrefmark{1}, Brian Jackson\IEEEauthorrefmark{1}}
\IEEEauthorblockA{\IEEEauthorrefmark{1}SRON,
Sorbonnelaan 2, 3584CA Utrecht, the Netherlands\\ Email: see http://www.sron.nl}
\IEEEauthorblockA{\IEEEauthorrefmark{2}VTT, Tietotie, Espoo, Finland}}


%

\IEEEspecialpapernotice{Accepted for publication in IEEE Transactions on Applied Superconductivity, DOI 10.1109/TASC.2015.2393716\\Copyright has been transferred to the IEEE.}

\maketitle

\begin{abstract}

TES-based bolometer and microcalorimeter arrays with thousands of pixels are under development for several space-based and ground-based applications. A linear detector response and low levels of cross talk facilitate the calibration of the instruments. In an effort to improve the properties of TES-based detectors, fixing the TES resistance in a resistance-locked loop (RLL) under optical loading has recently been proposed. Earlier theoretical work on this mode of operation has shown that the detector speed, linearity and dynamic range should improve with respect to voltage biased operation. 

This paper presents an experimental demonstration of multiplexed readout in this mode of operation in a TES-based detector array with noise equivalent power values (NEP) of $3.5\cdot 10^{-19} $W/$\sqrt{\mathrm{Hz}}$. The measured noise and dynamic properties of the detector in the RLL will be compared with the earlier modelling work. Furthermore, the practical implementation routes for future FDM systems for the readout of bolometer and microcalorimeter arrays will be discussed.

\end{abstract}



%
\IEEEpeerreviewmaketitle

\section{Introduction}

\begin{figure*}
\centering
\includegraphics[width=1.8\columnwidth]{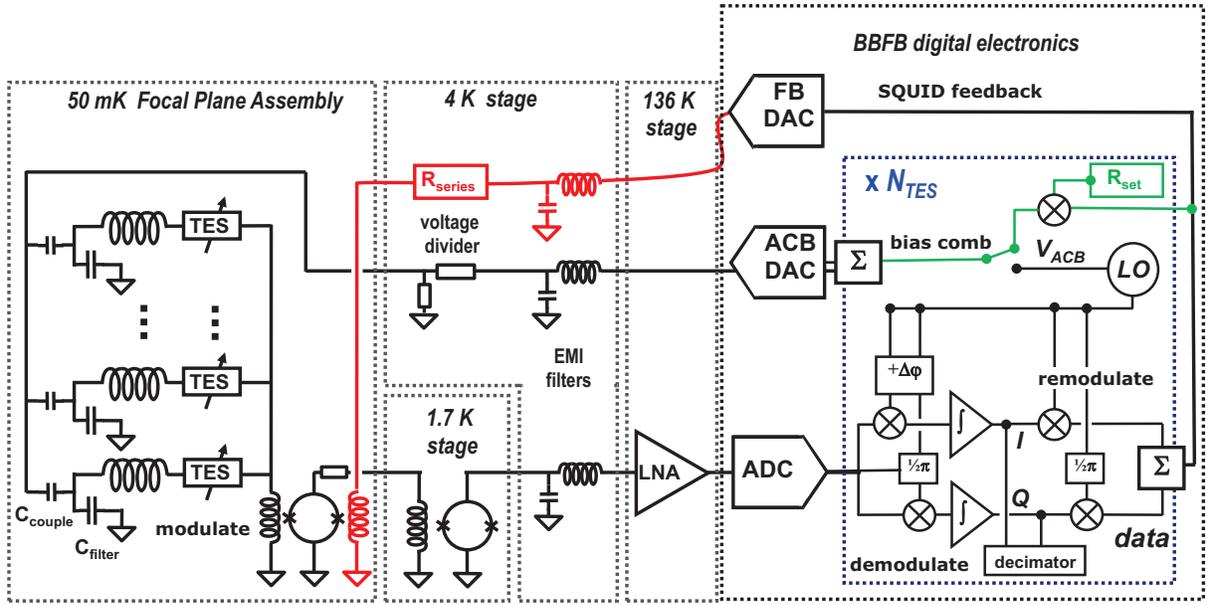}
\caption{Schematic diagram of the 8-pixel frequency domain multiplexed TES readout system. The resistance locked loop is integrated in the baseband feedback electronics as drawn on the right hand side. The envelope of the AC bias current is present in the $I$-channel. The AC bias voltage is equal to this measured current times the required set point resistance of the TES ($R_{\mathrm{set}}$). With the switch the user can choose between voltage bias and RLL operation. }
\label{fig:setup}
\end{figure*}

Imaging detector arrays based on Transition Edge Sensors (TES) are a maturing technology for a range of both earth and space-base applications, such as infrared astronomy, X-ray astronomy, and material analysis. SQUID-based multiplexed readout is needed to obtain a detector noise limited readout chain. Multiplexing in this context implies that multiple TES signals which overlap in time and phase space are made mathematically independent so that they can be transported through a shared readout chain without losing information, by multiplying each TES signal with an independent carrier and subsequently adding these modulated TES signals. 

The various multiplexing variants which are being developed can be ordered based on their modulating element. We distinguish the variants in which the SQUID is the multiplying element which mounts the signal on a carrier, and the variant in which the TES mounts the signal on the carrier (FDM). For the first category the TES is biased with a direct voltage (DC), and the modulated signals are separated in the time domain (TDM)\cite{irwin02}, or frequency domain (microwave SQUID multiplexing \cite{noroozian13} or code domain multiplexing \cite{niemack10}). For the second category the TES is biased with an alternating voltage (AC), and the modulated signals are separated in the frequency domain (FDM)\cite{dobbs12, kuur12}. 

For both variants of multiplexing the signals have to be bandwidth limited to prevent addition of wide-band noise by using an inductor as low-pass filter, or an inductor and capacitor in series as band-pass filter, respectively. An important difference between the two categories of multiplexing within the context of this paper, is the property that for case of SQUID-based modulation the bias sources of the TESes are commonly shared between pixels to lower the total wire count, with the consequence that the set points of the pixels cannot be chosen individually, whereas for the case of TES-based modulation the TES bias sources are independent and therefore multiplexed. This reduces the wire count and makes the working points of the pixels individually adjustable.

The individual bias setting per pixel can also be used to change the impedance of the TES bias source for each pixel by applying feedback to the applied bias voltage based on the measured TES current. This type of feedback has been proposed in different incarnations and for different purposes. It has been shown \cite{nam99} that with this type of feedback the effective thermal decay time of X-ray detectors can be made shorter when the effective internal resistance of bias source is made negative, i.e. $R_0< R_L<0$, where $R_0$ is the operating resistance of the TES, and $R_L$ the total series resistance in the TES bias circuit. The same feedback topology has also been applied to compensate parasitic resistances in the TES bias circuit \cite{dehaan12}, to increase the stiffness of the voltage bias and thereby the electro thermal feedback (ETF) loop gain, leaving a (slightly) net positive resistance $0<R_L\ll R_0$ in the voltage source.  Recently, the authors of this paper proposed \cite{kuur13} to use the regime of $R_L=-R_0$ for bolometer operation, so that the TES resistance is kept constant under optical loading, and that the small signal parameters of the TES change minimally under optical loading. As a result, the linearity of the detector is optimised, and  the dependence of the cross talk on the signal level is minimised. 

In this paper, we show the first experimental results which demonstrate that operation in the regime where $R_L=-R_0$ is stable both for single pixel and for multiplexed operation, and that the signal-to-noise ratio of the TES-based detector is conserved under the RLL with respect to voltage bias. For practical reasons the experiments have been divided over two setups. The comparison of the signal-to-noise ratio was performed in a bolometer pixel with a low NEP of $\sim 3.5 \cdot 10^{-19}$W/$\sqrt{\mathrm{Hz}}$. The multiplexing experiment was performed with X-ray micro calorimeters, because pulse-tube induced micro-vibrations allowed for stable operation of only one pixel in the bolometer multiplexer during this particular experiment. 

\section{Experimental setup}

\begin{figure}
\centering
 \includegraphics[width=0.8\columnwidth]{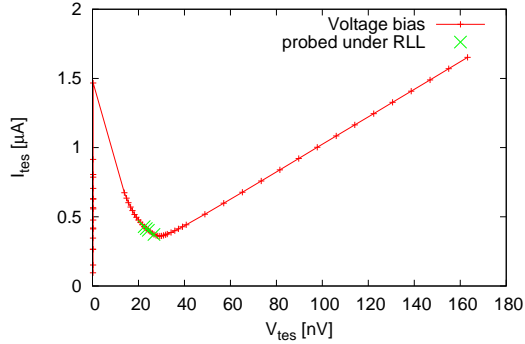}
 \caption{$IV$-curve of a bolometer pixel as taken under voltage bias. The pixel consists of a TES and absorber on a SiN island, which is suspended by four SiN legs of $400 \times 2 \times 0.25 \mu\mathrm{m}^3$. The size of the TiAu TES equals $50 \times 50 \mu$m, with $T_c= 80$~mK, a bias power level of $\sim 9$~fW, and a measured dark NEP of $\sim 3.5 \cdot 10^{-19}$W/$\sqrt{\mathrm{Hz}}$ \cite{gottardi14}. The bias points under which the pixel has been characterised using the RLL are indicated with the crosses. }
 \label{fig:ivcurve}
 \end{figure}

\begin{figure*}
\centering
\includegraphics[height=0.65\columnwidth]{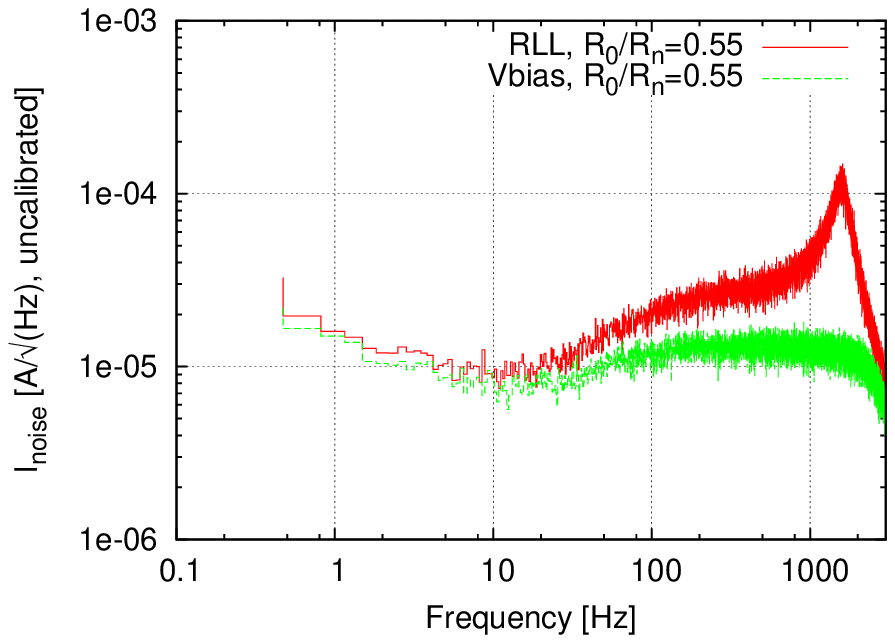}%
\includegraphics[height=0.65\columnwidth]{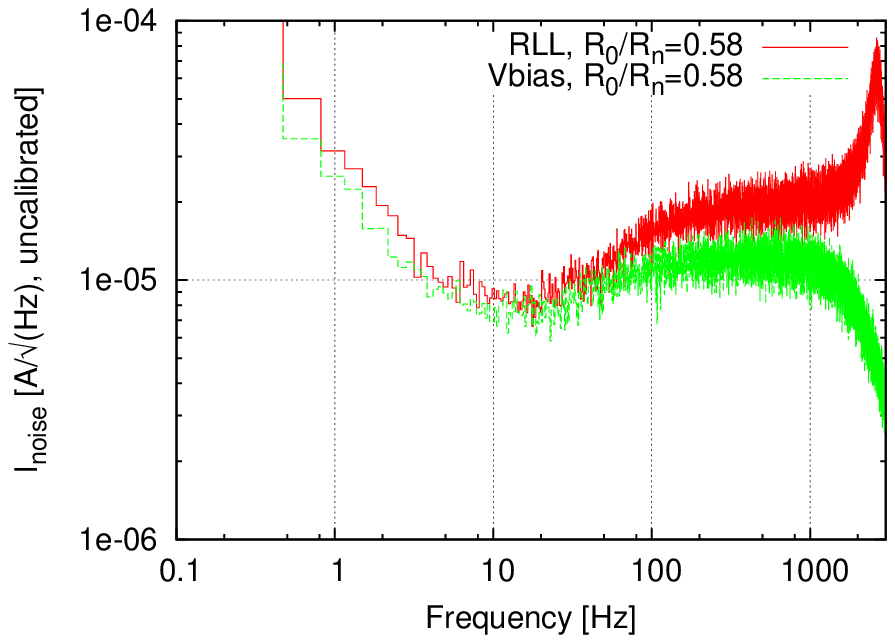}
\includegraphics[height=0.65\columnwidth]{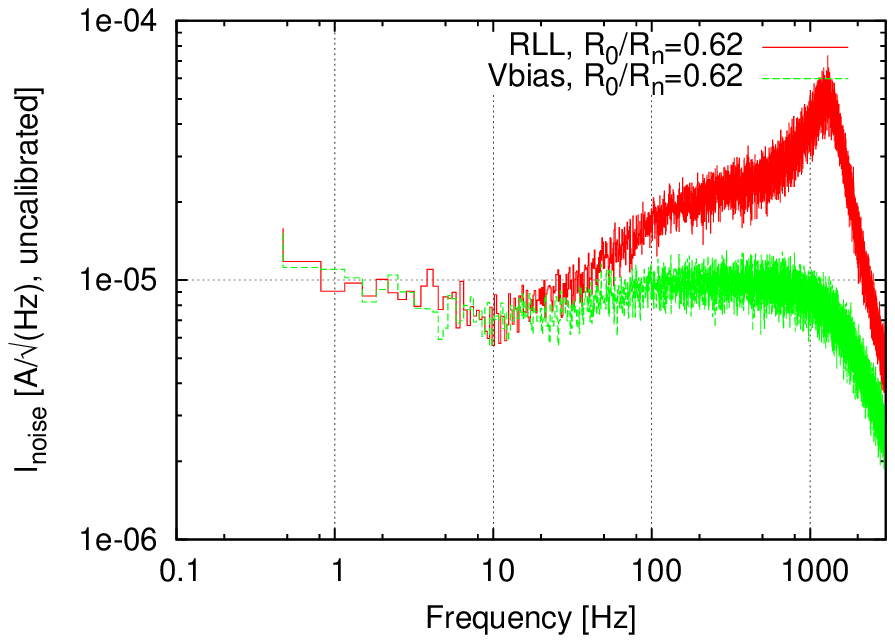}%
\includegraphics[height=0.65\columnwidth]{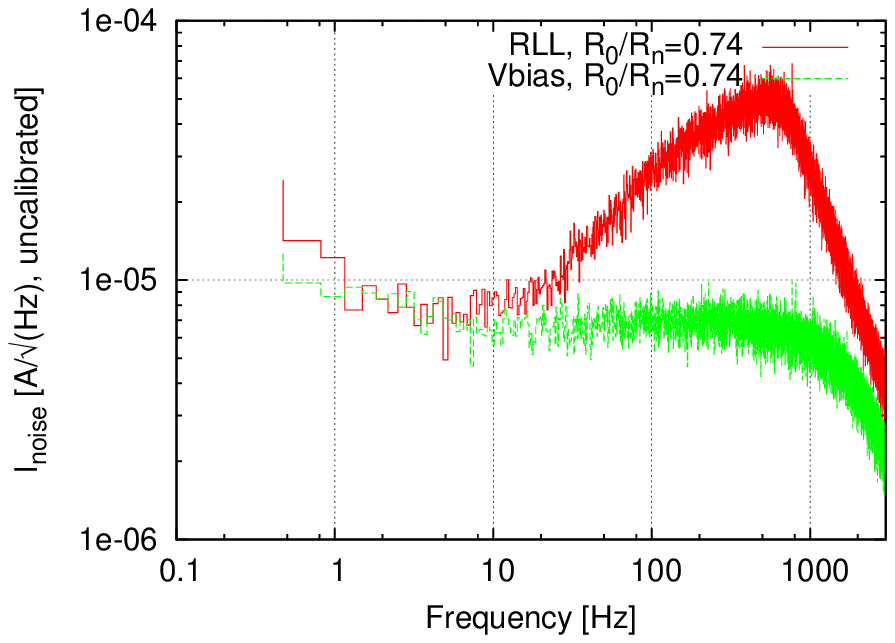}
\caption{Current noise spectral densities as measured for different bias points in the transitions, as observed under voltage bias and RLL.  Note that the spectra obtained under the RLL have been multiplied by a factor of 2, to make a direct comparison of the low-frequency end of the spectra possible. The information bandwidth equals the frequency at which the spectral density under the RLL is a factor of two higher than under voltage bias ($\sim 200$ Hz). In the low-frequency limit the observed current noise spectra overlap as expected.  For frequencies above the information bandwidth a higher noise level spectral density is observed under RLL, which is consistent with the expected behaviour for small values of $\beta$. The onset of electrothermal oscillations is observed in the two upper panels under RLL, which is a result of the decrease of the electrothermal decay time under the RLL. }
\label{fig:rll-vb-noise}
\end{figure*}

The experiments were performed with a 3-pixel FDM setup with integrated lithographic 8-pixel $LC$ bandpass filters \cite{bruijn14}, and a two-pixel X-ray FDM setup, both with baseband feedback (BBFB) electronics which provides SQUID linearisation and signal demodulation. The RLL was integrated in the firmware of the BBFB electronics, following the block diagram as sketched in Fig.~\ref{fig:setup}. The system is adjusted such that the dissipative TES current $I_R$ is aligned fully in the $I$-channel of the readout. The $Q$-channel feedback is required to ensure stability of the BBFB system. The output of the $I$-channel, i.e. the measured TES current, is multiplied with a constant  ($R_{\mathrm{set}}$), which is proportional to the resulting set point resistance of the TES, as the applied bias voltage $V_{bias}=I_R \cdot R_{\mathrm{set}}$. The switch toggles between the (standard) voltage bias condition and the RLL mode. 

\section{Experimental results}

To demonstrate the stability of the RLL, and to compare the signal-to-noise ratio of a bolometer pixel under voltage bias and the RLL, a typical bolometer pixel under development for the SW band of SAFARI \cite{jackson12} was used, at a bias frequency of 1.46~MHz. The pixel consists of a TES and absorber on a SiN island, which is suspended by four SiN legs of $400 \times 2 \times 0.25 \mu\mathrm{m}^3$. The size of the TiAu TES equals $50 \times 50 \mu$m, with $T_c= 80$~mK, a bias power level of $\sim 9$~fW, and a measured dark NEP of $\sim 3.5 \cdot 10^{-19}$W/$\sqrt{\mathrm{Hz}}$ \cite{gottardi14}. The pixel has been characterised for a range of bias points in the transition under both voltage bias and the RLL. The current-voltage characteristic of the pixel as obtained under voltage bias is shown in Fig.~\ref{fig:ivcurve}. The bias points which have been characterised using the RLL are marked with crosses. The latter range is limited because of electrothermal stability constraints, which originate from the much higher electrothermal loop gain under the RLL\cite{kuur13}, in combination with the fact that the circuit was originally designed for operation under voltage bias only. 

\subsection{Noise spectra}

Before discussing the results, we need to consider that under voltage bias the observed current noise spectra density and the noise equivalent power (NEP) scale linearly, as the bias voltage is kept constant. Under the RLL the situation is different, as not the bias voltage is kept constant, but the TES resistance instead. As a result, the detector NEP is proportional to the TES bias current $I(t)$ squared, as the dissipated power equals $P(t)=V(t)*I(t)=R_{\mathrm{set}}*I(t)^2$. In the small signal limit this implies that to calculate the NEP the observed current noise spectral density must be multiplied by 2, as the square of the observed current $I(t)$ equals $(I_0+\Delta I (t))^2 \approx I_0^2 + 2 I_0 \Delta I(t)$ in the small signal limit. 

From the RLL small signal model it follows that the detector NEP is equal under voltage bias and the RLL. This implies that also the information bandwidth of the detector, i.e. the frequency at which the phonon noise and Johnson noise of the detector are equal, is the same under RLL and voltage bias. It also follows from the small signal model that the measured current noise spectral densities under voltage bias and the RLL cross sect at the information bandwidth. For frequencies above the information bandwidth of the detector, but below the electrical bandwidth, the behaviours diverge. For that regime under voltage bias we expect to observe the current noise spectral density corresponding to a resistor of $R_0(1+\beta)$, whereas under the RLL this resistor value equals $R_0 \beta$. 

The measured current noise power spectral densities for four different bias points ($R_0/R_n=$ 0.55, 0.58, 0.62, and 0.74) in the transition are shown in Fig.~\ref{fig:rll-vb-noise}. To make a direct comparison with the noise power spectral density under voltage bias possible, the spectra as measured under the RLL have been multiplied by a factor of two, as discussed above. As a result of this multiplication, the information band is not found anymore at the cross section between the spectra, but at the frequency where the difference equals a factor of 2. Note that as we intend to do a relative comparison, and because we don't have an optical signal source available to measure the responsivity of the detector, the observed current noise spectral density has not been converted into calibrated NEP values. Instead, we present the current noise spectral densities.

It is clear that in the low frequency region, i.e. $f < 200$~Hz which is within the information bandwidth of the detector, the noise spectra overlap, as was to be expected as discussed above. We also observe that for higher frequencies the observed noise power is much higher for the RLL than under voltage bias. This qualitatively scales with the expected factor $1+1/\beta$, assuming that $\beta \ll1$. Unfortunately impedance data is lacking for this detector so that a quantitative comparison is not possible. However, the onset of electrothermal oscillations between 1 and 2~kHz shows that the thermal bandwidth of the detector, which is approximately equal to the information bandwidth ($\sim 200$~Hz),  has indeed shifted with a significant factor of $>5$ with respect to the situation under voltage bias.

\subsection{multiplexing results}

\begin{figure}
\includegraphics[width=\columnwidth]{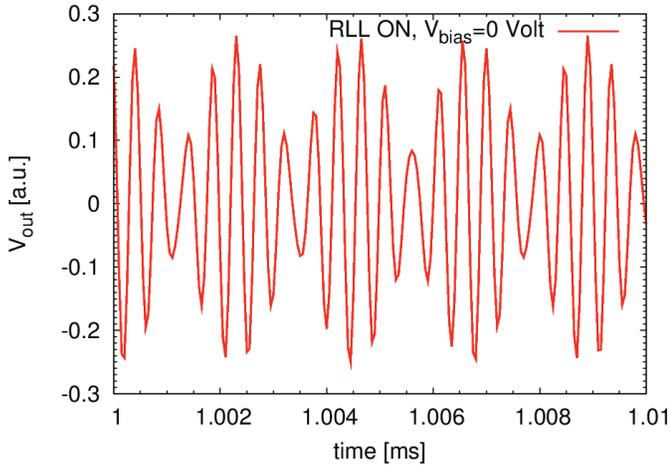}
\caption{Example of the beating in the bias currents at the output of the SQUID, which is observed when 2 bias currents run simultaneously under the RLL. Note that the observed currents are used for biasing the TESs, to close the RLL. The fact that the oscillation is stable shows that the RLL can be operated in more than 1 pixel simultaneously.}
\label{fig:2pix}
\end{figure}

To demonstrate that the RLL can also be applied for multiple pixels simultaneously, we put two X-ray microcalorimeter pixels simultaneously under resistance locking. It was found that the pixels operated stable, and that the observed dynamic behaviour and NEP did not change when the second pixel was added (not shown). Note that the observed currents are used for biasing the TESs, to close the RLL (see Fig.~\ref{fig:setup}). As an illustration of the resulting beating in the current is shown in the time domain plot in Fig.~\ref{fig:2pix}.  

\section{Conclusion}

Frequency domain multiplexing makes it possible to create a resistance locked loop per pixel, as the pixel bias voltages are multiplexed and therefore individually tuneable. The RLL improves detector speed, linearity and dynamic range for bolometer operation. It was shown that for FDM resistance locking can easily be integrated in the digital baseband feedback electronics which is used for demodulation and SQUID linearisation. It was found that within the information bandwidth of the detector the observed NEPs are equal, and that closing multiple loops simultaneously does not impair the stability of the system. When combined with the earlier observation \cite{gottardi14} that for TES-based bolometers the performance under AC bias is equal to under DC bias, we believe that this mode of operation is suitable for larger TES-based bolometer applications.

\bibliographystyle{IEEEtran}

\begin{thebibliography}{10}
\providecommand{\url}[1]{#1}
\csname url@samestyle\endcsname
\providecommand{\newblock}{\relax}
\providecommand{\bibinfo}[2]{#2}
\providecommand{\BIBentrySTDinterwordspacing}{\spaceskip=0pt\relax}
\providecommand{\BIBentryALTinterwordstretchfactor}{4}
\providecommand{\BIBentryALTinterwordspacing}{\spaceskip=\fontdimen2\font plus
\BIBentryALTinterwordstretchfactor\fontdimen3\font minus
  \fontdimen4\font\relax}
\providecommand{\BIBforeignlanguage}[2]{{%
\expandafter\ifx\csname l@#1\endcsname\relax
\typeout{** WARNING: IEEEtran.bst: No hyphenation pattern has been}%
\typeout{** loaded for the language `#1'. Using the pattern for}%
\typeout{** the default language instead.}%
\else
\language=\csname l@#1\endcsname
\fi
#2}}
\providecommand{\BIBdecl}{\relax}
\BIBdecl

\bibitem{irwin02}
K.~Irwin, ``{SQUID} multiplexers for transition-edge sensors,'' \emph{Physica
  C}, vol. 368, pp. 203--210, 2002.

\bibitem{noroozian13}
\BIBentryALTinterwordspacing
O.~Noroozian, J.~A.~B. Mates, D.~A. Bennett, J.~A. Brevik, J.~W. Fowler,
  J.~Gao, G.~C. Hilton, R.~D. Horansky, K.~D. Irwin, Z.~Kang, D.~R. Schmidt,
  L.~R. Vale, and J.~N. Ullom, ``High-resolution gamma-ray spectroscopy with a
  microwave-multiplexed transition-edge sensor array,'' \emph{Applied Physics
  Letters}, vol. 103, no.~20, pp.~--, 2013. [Online]. Available:
  \url{http://scitation.aip.org/content/aip/journal/apl/103/20/10.1063/1.4829156}
\BIBentrySTDinterwordspacing

\bibitem{niemack10}
M.~Niemack, J.~Beyer, H.~Cho, W.~Doriese, G.~Hilton, K.~Irwin, C.~Reintsema,
  D.~Schmidt, J.~Ullom, and L.~Vale, ``Code-division squid multiplexing,''
  \emph{Appl. Phys. Lett.}, vol.~96, p. 163509, 2010.

\bibitem{dobbs12}
\BIBentryALTinterwordspacing
M.~A. Dobbs, M.~Lueker, K.~A. Aird, A.~N. Bender, B.~A. Benson, L.~E. Bleem,
  J.~E. Carlstrom, C.~L. Chang, H.-M. Cho, J.~Clarke, T.~M. Crawford, A.~T.
  Crites, D.~I. Flanigan, T.~de~Haan, E.~M. George, N.~W. Halverson, W.~L.
  Holzapfel, J.~D. Hrubes, B.~R. Johnson, J.~Joseph, R.~Keisler, J.~Kennedy,
  Z.~Kermish, T.~M. Lanting, A.~T. Lee, E.~M. Leitch, D.~Luong-Van, J.~J.
  McMahon, J.~Mehl, S.~S. Meyer, T.~E. Montroy, S.~Padin, T.~Plagge, C.~Pryke,
  P.~L. Richards, J.~E. Ruhl, K.~K. Schaffer, D.~Schwan, E.~Shirokoff, H.~G.
  Spieler, Z.~Staniszewski, A.~A. Stark, K.~Vanderlinde, J.~D. Vieira, C.~Vu,
  B.~Westbrook, and R.~Williamson, ``Frequency multiplexed superconducting
  quantum interference device readout of large bolometer arrays for cosmic
  microwave background measurements,'' \emph{Review of Scientific Instruments},
  vol.~83, no.~7, pp.~--, 2012. [Online]. Available:
  \url{http://scitation.aip.org/content/aip/journal/rsi/83/7/10.1063/1.4737629}
\BIBentrySTDinterwordspacing

\bibitem{kuur12}
J.~van~der Kuur, J.~Beyer, M.~Bruijn, J.~Gao, R.~den Hartog, R.~Heijmering,
  H.~Hoevers, B.~Jackson, B.~van Leeuwen, M.~Lindeman, M.~Kiviranta,
  P.~de~Korte, P.~Mauskopf, P.~de~Korte, H.~van Weers, and S.~Withington, ``The
  spica-safari tes bolometer readout: Developments towards a flight system,''
  \emph{J. Low Temp. Phys.}, vol. 167, pp. 561--567, 2012.

\bibitem{nam99}
S.~W. Nam, B.~Cabrera, P.~Colling, R.~Clarke, E.~Figueroa-Feliciano, A.~Miller,
  and R.~Romani, ``A new biasing technique for transition edge sensor with
  electrothermal feedback,'' \emph{IEEE Trans. Appl. Supercond.}, vol.~9,
  no.~2, pp. 4209--4212, 1999.

\bibitem{dehaan12}
\BIBentryALTinterwordspacing
T.~de~Haan, G.~Smecher, and M.~Dobbs, ``Improved performance of tes bolometers
  using digital feedback,'' vol. 8452, 2012, pp. 84\,520E--84\,520E--10.
  [Online]. Available: \url{http://dx.doi.org/10.1117/12.925658}
\BIBentrySTDinterwordspacing

\bibitem{kuur13}
\BIBentryALTinterwordspacing
J.~van~der Kuur and M.~Kiviranta, ``Operation of transition edge sensors in a
  resistance locked loop,'' \emph{Applied Physics Letters}, vol. 102, no.~2, p.
  023505, 2013. [Online]. Available:
  \url{http://link.aip.org/link/?APL/102/023505/1}
\BIBentrySTDinterwordspacing

\bibitem{bruijn14}
\BIBentryALTinterwordspacing
M.~Bruijn, L.~Gottardi, R.~den Hartog, J.~van~der Kuur, A.~van~der Linden, and
  B.~Jackson, ``\BIBforeignlanguage{English}{Tailoring the high-q lc filter
  arrays for readout of kilo-pixel tes arrays in the spica-safari
  instrument},'' \emph{\BIBforeignlanguage{English}{Journal of Low Temperature
  Physics}}, vol. 176, no. 3-4, pp. 421--425, 2014. [Online]. Available:
  \url{http://dx.doi.org/10.1007/s10909-013-1003-6}
\BIBentrySTDinterwordspacing

\bibitem{jackson12}
B.~D. Jackson, P.~A.~J. de~Korte, J.~van~der Kuur, P.~D. Mauskopf, J.~Beyer,
  M.~P. Bruijn, A.~Cros, J.-R. Gao, D.~Griffin, R.~den Hartog, M.~Kiviranta,
  G.~de~Lange, B.-J. van Leeuwen, C.~Macculi, L.~Ravera, N.~Trappe, H.~van
  Weers, and S.~Withington, ``The spica-safari detector system: Tes detector
  arrays with frequency division multiplexed squid readout,'' \emph{IEEE Trans.
  THz Sci. Tech.}, vol.~2, no.~1, pp. 12--21, 2012.

\bibitem{gottardi14}
L.~{Gottardi}, H.~{Akamatsu}, M.~{Bruijn}, J.-R. {Gao}, R.~{den Hartog},
  R.~{Hijmering}, H.~{Hoevers}, P.~{Khosropanah}, A.~{Kozorezov}, J.~{van der
  Kuur}, A.~{van der Linden}, and M.~{Ridder}, ``{Weak-Link Phenomena in
  AC-Biased Transition Edge Sensors},'' \emph{Journal of Low Temperature
  Physics}, vol. 176, pp. 279--284, Aug. 2014.

\end{thebibliography}

\end{document}